\documentclass[proof]{pasj01}
\draft

\Received{}
\Accepted{}
 
\begin{document} 

\title{ 
Extinction and dust/gas ratio in the H\,\emissiontype{I} ridge region of the LMC based on the IRSF/SIRIUS near-infrared survey}

\author{Takuya \textsc{Furuta}\altaffilmark{1}$^{*}$}%
\altaffiltext{1}{Graduate School of Science, Nagoya University, Furo-cho, Chikusa-ku, Nagoya, Aichi 464-8602, Japan}
\email{t.furuta@u.phys.nagoya-u.ac.jp}

\author{Hidehiro \textsc{Kaneda}\altaffilmark{1}}%

\author{Takuma \textsc{Kokusho}\altaffilmark{1}}%

\author{Daisuke \textsc{Ishihara}\altaffilmark{1}}%

\author{Yasushi \textsc{Nakajima}\altaffilmark{2}}%
\altaffiltext{2}{Center for Information and Communication Technology, Hitotsubashi University, 2-1 Naka, Kunitachi, Tokyo 186-8601, Japan}

\author{Yasuo \textsc{Fukui},\altaffilmark{1}}

\author{Kisetsu \textsc{Tsuge},\altaffilmark{1}}


\maketitle

\begin{abstract}
We present a dust extinction $A_V$ map of the Large Magellanic Cloud (LMC) in the H\,\emissiontype{I} ridge region using the IRSF near-infrared (IR) data, and compare the $A_V$ map with the total hydrogen column density $N$(H) maps derived from the CO and H\,\emissiontype{I} observations.
In the LMC H\,\emissiontype{I} ridge region, the two-velocity H\,\emissiontype{I} components (plus an intermediate velocity component) are identified, and the young massive star cluster is possibly formed by collison between them.
In addition, one of the components is suggested to be an inflow gas from the Small Magellanic Cloud (SMC) which is expected to have even lower metallicity gas (Fukui et al. 2017, PASJ, 69, L5).
To evaluate dust/gas ratios in the H\,\emissiontype{I} ridge region in detail, we derive the $A_V$ map from the near-IR color excess of the IRSF data updated with the latest calibration, and fit the resultant $A_V$ map with a combination of the $N$(H) maps of the different velocity components to successfully decompose it into the 3 components.
As a result, we find difference by a factor of 2 in $A_{V}$/$N$(H) between the components.
In additon, the CO-to-H$_2$ conversion factor also indicates difference between the components, implying the difference in the metallicity.
Our results are likely to support the scenario that the gas in the LMC H\,\emissiontype{I} ridge region is contaminated with an inflow gas from the SMC with a geometry consistent with the on-going collision between the two velocity components.
\end{abstract}

\section{Introduction}
The Large Magellanic Cloud (LMC) is the galaxy closest to us, which enables us to perform spatially well-resolved studies of the interstellar medium (ISM) in external galaxies.
The LMC is also known to be a nearly face-on galaxy ($i\sim35^{\circ}$; \cite{inclination}), allowing us to discuss stellar and gas distributions in almost two dimensions.
In addition, the LMC has a low-metallicity, $Z \sim$ 0.3--0.5 $\rm Z_{\odot}$ (\cite{metal}), a typical value of the ISM at redshift $z\sim1.5$ (\cite{redshift}), around which epoch the Universe experienced the most active phase of star formation in galaxies.
Thanks to these characteristics, the LMC is an excellent laboratory for studying star formation and evolution in low-metallicity environments.

A recent study suggested that the formation of the young massive cluster R136 in the LMC was triggered by the collision of H\,\emissiontype{I} clouds (\cite{fukui_2017}). 
They identified two-velocity H\,\emissiontype{I} components and found the bridge features connecting the two components, which is the evidence for the collision between them.  
They also compared the H\,\emissiontype{I} intensity with the optical depth of the dust emission derived from the Planck/IRAS data, and found that the ratio of the H\,\emissiontype{I} intensity to the dust optical depth in the H\,\emissiontype{I} ridge region is different from that outside the H\,\emissiontype{I} ridge region.
They suggested that the gas in the LMC H\,\emissiontype{I} ridge region may be mixed with an inflow gas from the Small Magellanic Cloud (SMC) which is known to have even lower metallicity gas, based on the result of the numerical simulation (\cite{shock_sim}). 
Hence, the massive star formation in the LMC is possibly triggered by the interaction between the galaxies, and thus it is important to investigate the dust/gas ratios in the LMC to prove this scenario.
As a method to estimate the dust abundance, the interstellar dust extinction is useful, because it does not depend much on physical conditions of dust, such as the dust temperature, although it has no information on the velocity of the ISM. 

The methods of measuring the interstellar dust extinction have so far been developed by many authors in the past.
The near-infrared (IR) color excess (NICE) method suggested by \citet{nice} derived the dust extinction with the $H-K$ color excess and was proven  to be a valid method of mapping the dust extinction.
The NICER (NICE revised) method developed by \citet{nicer} combined the $J-H$ and $H-K$ colors for deriving the dust extinction.
Furthermore, ``$X$ percentile method'' is suggested by \citet{dobashi}.
In this method, the color excess was estimated more precisely than in the NICE and NICER methods owing to elimination of  galactic foreground stars and intrinsically red stars such as young stellar objects (YSOs) and asymptotic giant branch (AGB) stars.
By using these methods, the extinction maps of the LMC were constructed on the basis of the Two Micron All Sky Survey (2MASS) point source catalog (e.g., \cite{extinction_nice}; \cite{dobashi}).

In this paper, we update the dust extinction map of the LMC in the H\,\emissiontype{I} ridge region to discuss the dust distribution in the multiple cloud components of different velocities revealed by CO and H\,\emissiontype{I} observations (\cite{comap}; \cite{himap}).
 

\section{The data and the derivation of the extinction map}
\subsection{IRSF data}
\citet{irsf} presented a near-IR ($J,\  H,$ and $K_{\rm S}$ bands) photometric catalog for
the Magellanic Clouds obtained with the SIRIUS camera on the InfraRed Survey Facility (IRSF) 1.4 m telescope at the South African Astronomical Observatory (hereafter the IRSF catalog), the 10 $\sigma$ limiting magnitudes of which are 18.8, 17.8, and 16.6 mag in the $J,\  H,$ and $K_{\rm S}$ bands, respectively. 
As the limiting magnitudes of the IRSF catalog are about 3 magnitudes deeper than those of the 2MASS point source catalog, the stellar number density is expected to be about 6 times higher in the LMC.
Therefore, we adopted the IRSF catalog to make an extinction map toward the LMC with a finer resolution than the previous dust extinction maps from the near-IR color excess presented by \citet{extinction_nice} and \citet{dobashi}.

However, we found a systematic photometric error in the existing IRSF catalog; an extinction map using the IRSF catalog shows a grid structure that delineates the observed frames. 
Comparing the $J,\  H,$ and $K_{\rm S}$ bands magnitudes of the IRSF sources with the 2MASS $J,\  H,$ and $K_{\rm S}$ bands magnitudes, we consider that the flat-field correlation is likely inaccurate for the IRSF catalog, because the magnitude difference maps on the pixel coordinates show a clear increase near every edge of the array for all the bands.
This is presumably due to the reflection of the incident twilight sky illumination inside the IRSF/SIRIUS camera, since the reflection violates the assumption that the twilight sky illumination is uniform.

To solve this problem, we made a large grid on the magnitude difference map and computed the median for each grid.
As a result, we obtained a flux bias pattern with a maximum of 0.05 mag for each band. 
The flat-field image was then corrected with the resultant flux bias pattern for each band. 
We re-reduced the raw data of the IRSF Magellanic Cloud survey with the corrected flat-fielding images and updated the $J,\  H,$ and $K_{\rm S}$ bands magnitudes of the catalog.
We selected data from the catalog on the basis of the following criteria: (1) the photometric uncertainty is smaller than $0.11\ \rm mag$, corresponding to S/N$\geq10$ in all the bands; (2) the ``quality flag'' in the original catalog describing the shape of sources is 1 in all the bands, representing ``point-like'';
 and (3) the number of combined dithered images is larger than 8 in all the bands.

\subsection{Derivation of the dust extinction map}
We estimate the dust extinction on the basis of the NICER method (\cite{nicer}).
In our study, we estimate the color excess by measuring the difference between the observed color and the intrinsic color derived from \citet{tr_giant} on a color-color diagram (CC diagram).
These intrinsic and observed colors are converted into the Johnson system according to the conversion equations suggested by \citet{color_corr} and \citet{color_corr2}.
Hereafter, we present all the magnitudes and colors in the Johnson system.
\citet{dobashi} pointed out that the reddening law given by \citet{red_law2} represents the observed colors in the LMC better than that given by \citet{red_law1} which is adopted in the NICER method.
Therefore, we obtain the color excess assuming the reddening law by \citet{red_law2} which is determined by the near-IR photometry in the Johnson system.

\begin{figure}
 \begin{center}
  \includegraphics[width=8cm]{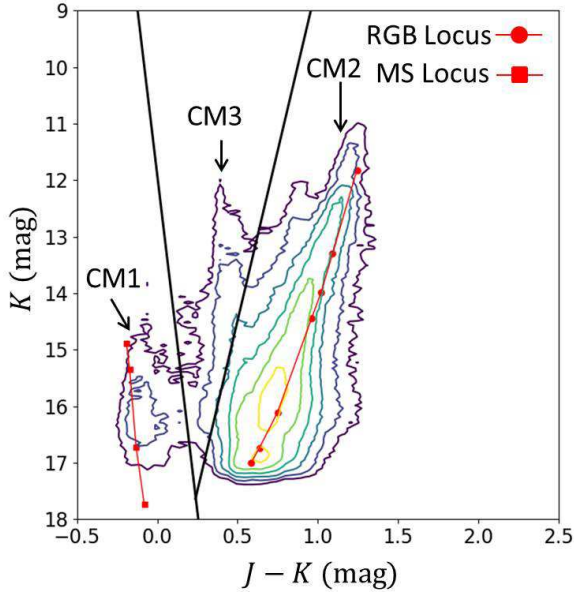} 
 \end{center}
\caption{Color-magnitude ($J-K$ vs. $K$) diagram of the selected sources of the LMC in the H\,\emissiontype{I} ridge region. Contours correspond to the number densities binned by 0.05 mag and its levels are logarithmically spaced from $10^{1.5}$ to $10^4$ with a step of $10^{0.5}$. The lines connected with the red circles and squares are the loci of the RGB and MS, respectively. Black solid lines indicate the boundary of each component (``CM1'' to ``CM3'').}\label{fig:color-mag}
\end{figure}

We need to eliminate foreground stars contaminating the selected sources, because the color excess is estimated by assuming that all the stars are located behind dark clouds. 
In order to eliminate such foreground stars, we first identify stellar populations by a color-magnitude diagram (CM diagram).
Figure \ref{fig:color-mag} shows the CM diagram ($J-K$ vs. $K$) in our selected data.
The loci of the main sequence (MS) and the red giant branch (RGB) stars in the LMC at the distance moduli of 18.5 (\cite{distance}) are also plotted in this diagram, where we can recognize three structural components defined as  ``CM1'' to ``CM3'' by \citet{irsf}.
\citet{2mass_pop} identified stellar populations, using the CM diagram of the LMC data from the 2MASS point source catalog.
Comparing ``CM1'' to ``CM3'' with the 12 components designated as ``A'' to ``L''  by \citet{2mass_pop}, \citet{irsf} suggested that ``CM1'', ``CM2'', and ``CM3'' primarily indicate MS (corresponding to ``A''), RGB (``E'', ``F'', ``G''), and Galactic forground stars (``B''), respectively.
Following the procedure by \citet{2mass_pop}, we define regions ``CM1'' to ``CM3'' in the CM diagram in figure \ref{fig:color-mag}.
In order to determine the boundary between ``CM2'' and ``CM3'', we fitted a linear function to the valley positions of the $J - K$ color histogram binned with $K=0.05 \ \rm mag$ at $K=12.0$ mag to $14.0$ mag, and extrapolated the fitted line to $K > 14.0$ mag.
We performed the same procedure at $K=14.5$ mag to $17.1$ mag to determine the boundary between ``CM1'' and ``CM3''.
The obtained boundary lines are shown in figure \ref{fig:color-mag}.
We use only the objects classified as RGB stars in the ``CM2'' region from the CM diagram to estimate the color excess.

\begin{figure}
 \begin{center}
  \includegraphics[width=17cm]{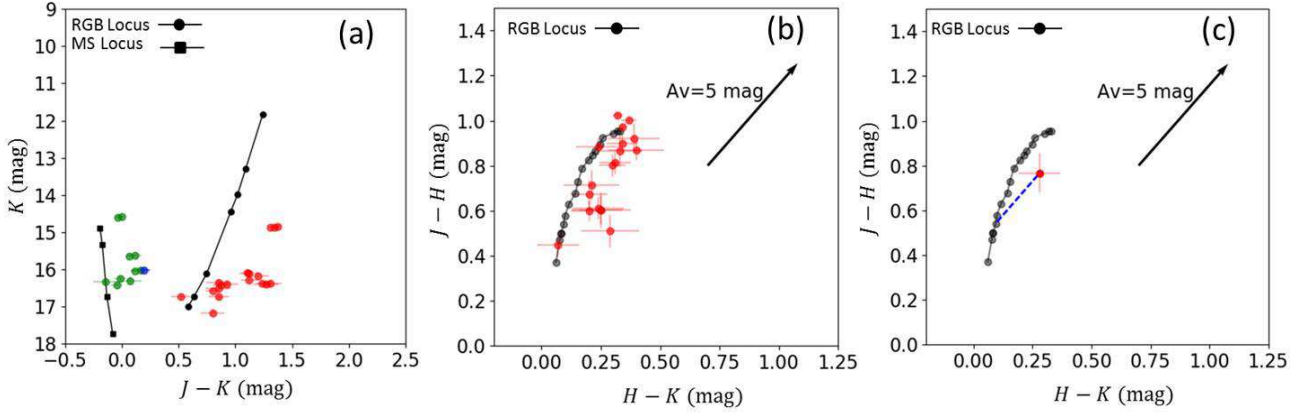} 
 \end{center}
\caption{Examples of the color-magnitude and color-color diagrams of stars within a spatial bin of \timeform{0'.5}. Black circles and squares are the loci of the RGB and MS, respectively. (a) Color-magnitude diagram of $J-K$ vs. $K$. Red, blue, and green circles are RGB, foreground stars, and MS, respectively, which are classified according to the boundary in figure \ref{fig:color-mag}.
(b) Color-color diagram of $H - K$ vs. $J - H$. Black arrow shows the reddening vector whose slope is 1.2.
(c) Color-color diagram of $H - K$ vs. $J - H$, where a red circle is the centroid position of the RGB colors. The blue dashed line is the reddening vector from the centroid position to the intrinsic color of RGB stars, corresponding to $A_V$ of 2.46 mag in this case.}\label{example}
\end{figure}

We calculate the dust extinction for each spatial bin with the resolution of \timeform{0'.5}.
We take the following three steps: first, on the basis of the classification of stellar populations by the CM diagram (figure \ref{example}a), we make the CC diagram for each spatial bin (figure \ref{example}b).
In this diagram, dusty AGB stars which possess intrinsically red colors ($J - K > 1.5$ mag and $H - K > 0.4$ mag; \cite{agb_color}) are included.
Contamination of such stars may prevent us from deriving a precise dust extinction.
Therefore, we perform $3\sigma$-clipping to both $J - H$ and $H - K$ colors, where $\sigma_{J-H}$ and $\sigma_{H-K}$ are defined as the standard deviations of the $J - H$ and $H - K$ colors, respectively, of the stars every spatial bin.

Second, we calculate the centroid positions of the RGB stars falling into a spatial bin and its color excess by measuring the difference between the centroid positions of RGB stars and the intrinsic color derived from \citet{tr_giant} on the CC diagram as shown in figure \ref{example}c.
Here, it should be noted that the color excess thus estimated to the LMC would be more or less underestimated unless all the RGB stars are located in the background of the clouds.
However, that geometry is expected in some particular areas of the H\,\emissiontype{I} ridge region, as will be discussed later.

Assuming the reddening law suggested by \citet{red_law2}, we have
\begin{equation}
A_{V}=10.9E(J - H)
\label{eq:excess1}
\end{equation}
and
 \begin{equation}
A_{V}=13.2E(H - K),
\label{eq:excess2}
\end{equation}
where $E(J - H)$ and $E(H - K)$ are the color excess of the $J-H$ and $H - K$ colors, respectively.
From equations (\ref{eq:excess1}) and (\ref{eq:excess2}), $E(J - H)/E(H - K)$ is estimated to be 1.2.
Thus, along with the reddening vector with a slope of 1.2, we finally estimate the dust extinction from the separation length between the intrinsic and observed colors of RGB stars on the CC diagram.

When there are no available RGB stars or intersection between the reddening vector and the intrinsic color of RGB stars in a spatial bin, ``not a number'' is assigned to the corresponding bin.
In the final procedure, we apply a median filter with the kernel size of \timeform{1'.6} to the dust extinction map so that the bins where ``not a number'' is assigned are given the median of the surrounding bins. 
\begin{figure}
 \begin{center}
  \includegraphics[width=9cm]{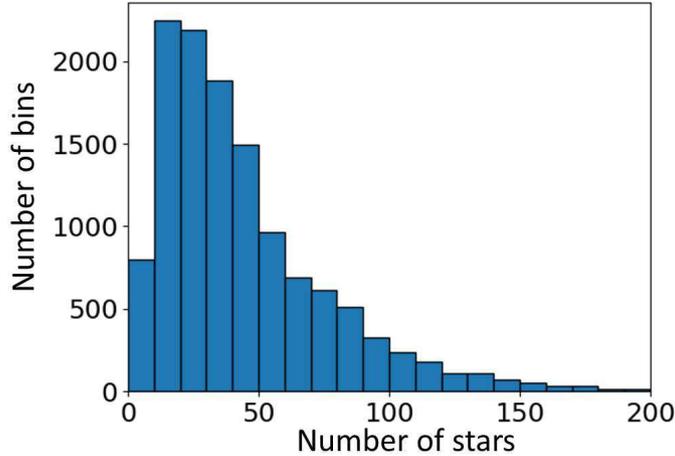} 
 \end{center}
\caption{Histogram of the number of the RGB stars included in each pixel of \timeform{1'.6}$\times$\timeform{1'.6}.}\label{fig:star_density}
\end{figure}
Figure \ref{fig:star_density} shows the histogram of the number of the RGB stars, $N$, included in each pixel of \timeform{1'.6}$\times$\timeform{1'.6}, where the averaged star density is 44.
We calculate the uncertainties of $A_V$, using the data in each individual pixel.
Based on the observed $\sigma_{J-H}$ and $\sigma_{H-K}$ with $N$ stars, we perform a Monte-Carlo simulation to estimate the error of $A_V$ per pixel, $\delta A_V$, by repeating the simulation 100 times under the assumption that the star colors follow the 2-dimensional Gaussian distribution in the CC diagram.
\section{Result}
\subsection{Extinction map}
\begin{figure}
 \begin{center}
  \includegraphics[width=16cm]{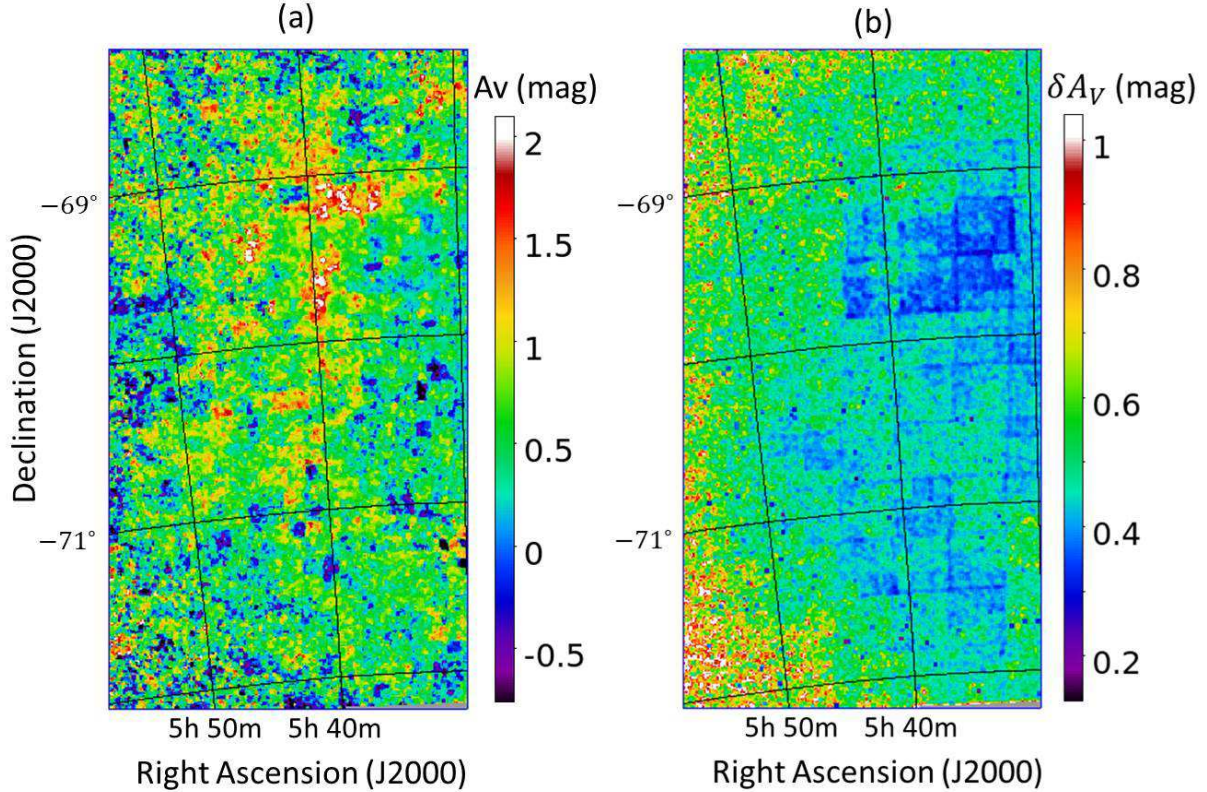} 
 \end{center}
\caption{(a) Extinction map of the LMC in the H\,\emissiontype{I} ridge region. The angular resolution of the map is \timeform{1.'6}. Color levels are given in units of $A_V$. (b) Map of the uncertainties of (a).}\label{fig:extinction}
\end{figure}

Figure \ref{fig:extinction}a shows the dust extinction map of the LMC obtained for the H\,\emissiontype{I} ridge region, while figure \ref{fig:extinction}b is the map of the uncertainties of the dust extinction.
In our extinction map, the mean visual extinction ($A_{V}$) is 0.53 mag and the mean noise level ($\delta A_V$) is 0.51 mag.
The negative extinction in the map means that the colors of stars in the spatial bin are bluer than the intrinsic color due to the photometric errors.
The region having high $A_{V}$ ($>$2.0 mag) corresponds to the 30 Dor region at $(\alpha, \delta)_{\rm J2000.0} =$ (\timeform{5h40m}, $-$\timeform{69D5'}).
From the dust extinction map, we can also recognize other well-known clouds such as ``molecular ridge'' at $\alpha_{\rm J2000.0} \sim $\timeform{5h40m} and $\delta_{\rm J2000.0} \sim$ \timeform{69D30'} to \timeform{71D} and  ``CO Arc'' identified by \citet{co_arc} at $(\alpha, \delta)_{\rm J2000.0} =$ (\timeform{5h44m}, $-$\timeform{69D30'}).

\begin{figure}
 \begin{center}
  \includegraphics[width=14cm]{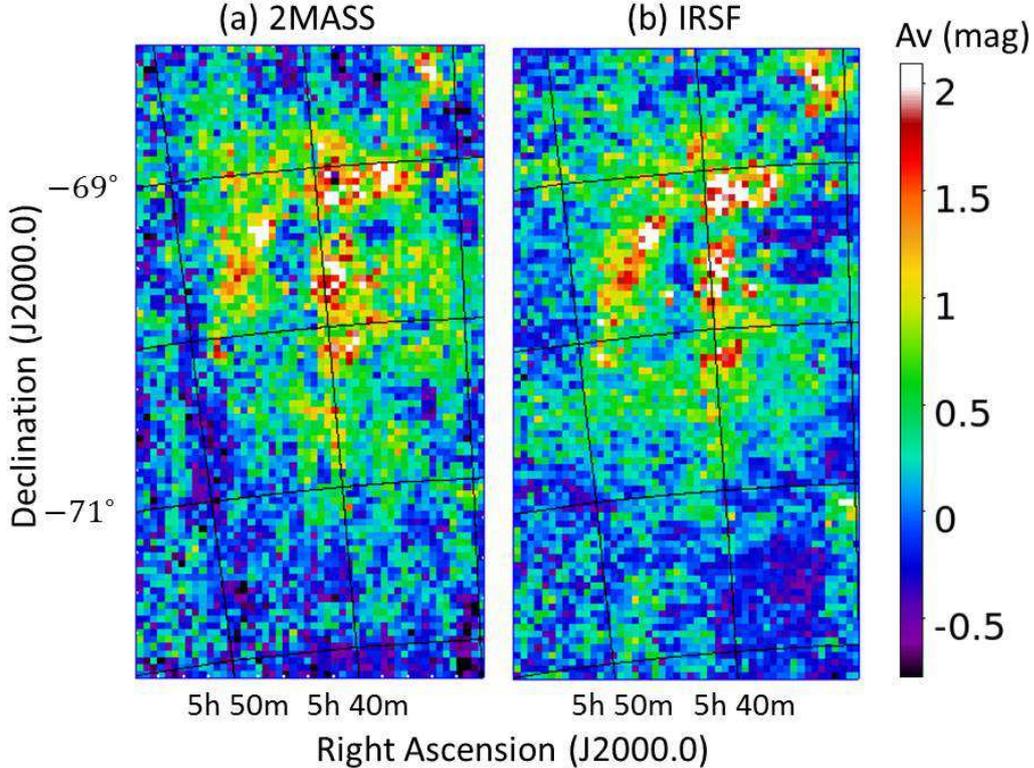} 
 \end{center}
\caption{(a) Dust extinction map of \citet{dobashi} with the 2MASS catalog and (b) that derived using the same method as in \citet{dobashi} with the IRSF catalog. Both maps cover the same region as in figure \ref{fig:extinction}, with the angular resolution of \timeform{2.'6}. Color levels are the same as those in figure \ref{fig:extinction}.}\label{fig:diff_2mass}
\end{figure}

We compare the dust extinction map with that derived by \citet{dobashi}.
Figure \ref{fig:diff_2mass}a shows the dust extinction map of \citet{dobashi} using the 2MASS catalog for the same region.
As can be seen in the figures, although the global distributions are similar, we can recognize local systematic differences in the molecular ridge and CO Arc regions.
The differences are most probably caused by the difference in the method of deriving $A_V$; we consider the intrinsic colors of several spectral types of the RGB stars, while \citet{dobashi} used one intrinsic color in their reference field.
In the latter case, late-type stars such as M giants can accidentally be regarded as reddened stars.
On the other hand, \citet{dobashi} managed to remove stars located on the near side of extinction sources with the $X$ percentile method selecting the stars in the range [$X_0$, $X_1$] percentile, while we do not apply the $X$ percentile method due to insufficient statistics in the numbers of the RGB stars included in a smaller bin size (section 2.2).
In order to verify this, we create the dust extinction map following the same method as in \citet{dobashi} with the range [$X_0$, $X_1$]=[80, 95]\% using the IRSF catalog.
The resultant map is displayed in figure \ref{fig:diff_2mass}b, which indeed shows an excellent agreement with figure \ref{fig:diff_2mass}a.
A comparison is also made on the errors of $A_V$ between our $A_V$ map (figure \ref{fig:extinction}) and that from \citet{dobashi}.
After re-gridding our $A_V$ map to the same angular resolution as theirs (\timeform{2'.6}$\times$\timeform{2'.6}), the errors in our $A_V$ map are found to distribute in a range of 0.2--0.5 mag, while those in \citet{dobashi} are reported to be 0.4--0.7 mag.
The improvement is smaller than expected from the fact that the limiting magnitude of IRSF is 3 magnitudes deeper than the 2MASS data used by \citet{dobashi}.
However, this will be explained by the difference in the method of deriving $A_V$, again; we use only RGB stars, while they used all the stars with the $X$ percentile method.
In the following result and discussion, we use the dust extinction map shown in figure \ref{fig:extinction}, because our map is less affected by the assumption of the intrinsic color.
As will be discussed later, a significant fraction of the clouds associated with the H\,\emissiontype{I} ridge region is likely to be in front of the LMC disk so that we expect that the $X$ percentile method would not change our conclusion significantly. 

\citet{fir_map} created the $A_V$ map from the Herschel far-IR dust continuum data and compared it with that from \citet{dobashi}.
They found that $A_V$ in the Herschel $A_V$ map is about 1.6 times higher than that from \citet{dobashi} systematically with a large scatter, although an overall spatial distribution is similar to each other.
They concluded that the large scatter is likely to be caused by the local geometry of the gas.
Comparing our $A_V$ map with the Herschel $A_V$ map, we also confirm that the result is almost the same as the previous study; an overall structure is quite similar, while some local differences are seen, for example, in the ``CO Arc'' region around $(\alpha, \delta)_{\rm J2000.0} =$ (\timeform{5h46m}, $-$\timeform{69D40'}).
In principle, far-IR-based $A_V$ maps trace the total dust column densities along the lines of sight, while near-IR-based $A_V$ maps selectively trace the dust column densities only in front of the background stars, and thus they would provide different information on the dust distribution complementary to each other.
As will be shown later, the feature of the near-IR-based $A_V$ maps is particularly useful for the geometry suggested by \citet{fukui_2017}.

 \subsection{$A_{V}$ vs. $N$(H) correlation}
 \subsubsection{Gas tracer}
As an atomic gas tracer, we use the H\,\emissiontype{I} velocity-integrated intensity maps derived by \citet{fukui_2017}, where the Australia Telescope Compact Array (ATCA) and Parkes H\,\emissiontype{I} $21\ \rm cm$ data (\cite{himap}) are used.
The angular resolution of the H\,\emissiontype{I} map is  \timeform{1.'0}, which corresponds to $\sim 15\ \rm pc$ for the LMC.
The rms noise level is $2.4\ \rm K$ with a velocity resolution of $1.649\ \rm km\ s^{-1}$.
\citet{fukui_2017} separated the H\,\emissiontype{I} map into two H\,\emissiontype{I} velocity components (for detail of the analysis, see \cite{tsuge}).
One is the H\,\emissiontype{I} velocity component existing over the whole disk of the LMC defined as D-component.
They calculated the relative velocity to the D-component, $V_{\rm offset}$, subtracting the galactic rotation velocity.
The velocity range of the D-component is $V_{\rm offset}\ =$ $-$10 to 10 $\rm km\ s^{-1}$. 
The other velocity component is spatially confined H\,\emissiontype{I} gas having velocity lower than the D-component with the integrated velocity range of $V_{\rm offset}=$ $-$100 to $-$30 $\rm km\ s^{-1}$.
They defined this component as L-component and suggested that the L-component may be of the origin of an inflow gas from the SMC.
In addition to these two components, we use the H\,\emissiontype{I} gas having the intermediate velocity between the L- and D-components ($V_{\rm offset}=$ $-$30 to $-$10 $\rm km\ s^{-1}$) defined as I-component by \citet{tsuge}
 
As a molecular gas tracer, we use the rotational transitions of $^{12}{\rm CO}\ (J=$1--0) observed with the NANTEN 4 $\rm m$ telescope located at Las Campanas Observatory in Chile (\cite{comap}).
The half-power beam width of the telescope is \timeform{2'.6}, and the velocity resolution is $0.65\ \rm km\ s^{-1}$.
The CO integrated intensity map is separated into L-, I-, and D-components according to the above velocity ranges of the H\,\emissiontype{I} integrated intensity maps.

We convert the H\,\emissiontype{I} integrated intensity into the H\,\emissiontype{I} column density by using the conversion factor, $X_{\rm H\,\emissiontype{I}}=1.82 \times 10^{18}\ \rm{cm^{-2}}/({\rm K\ km\ s^{-1}})$ (\cite{nh_conv}). 
CO can be used to trace $\rm H_{2}$; the CO-to-$\rm H_{2}$ conversion factor, $X_{\rm CO}$, can be obtained by assuming that the molecular clouds are gravitationally in equilibrium.
To estimate the $\rm H_{2}$ column density, we adopt $X_{\rm CO}=7\times10^{20}\ {\rm cm^{-2}}/(\rm{K\ km\ s^{-1}})$ which is the averaged  CO-to-$\rm H_{2}$ conversion factor in the LMC derived by \citet{h2_conv}.
We reduce the spatial resolution of the H\,\emissiontype{I} column density maps to be the same as that of the CO maps at a resolution of \timeform{2'.6}, and calculate the total gas column density $N({\rm H})$ as
\begin{equation}
N({\rm H})\ =\ N({\rm H\,\emissiontype{I}})\ + \ 2N({\rm H_{2}}),
\label{nh}
\end{equation}
where $N({\rm H\,\emissiontype{I}})$ and $N({\rm H_2})$ are the H\,\emissiontype{I} and the $\rm H_{2}$ column densities, respectively.
We calculate $N(\rm H)$ of each of the L-, I-, and D-components.

\subsubsection{Correlation results}
\begin{figure}
 \begin{center}
  \includegraphics[width=16cm]{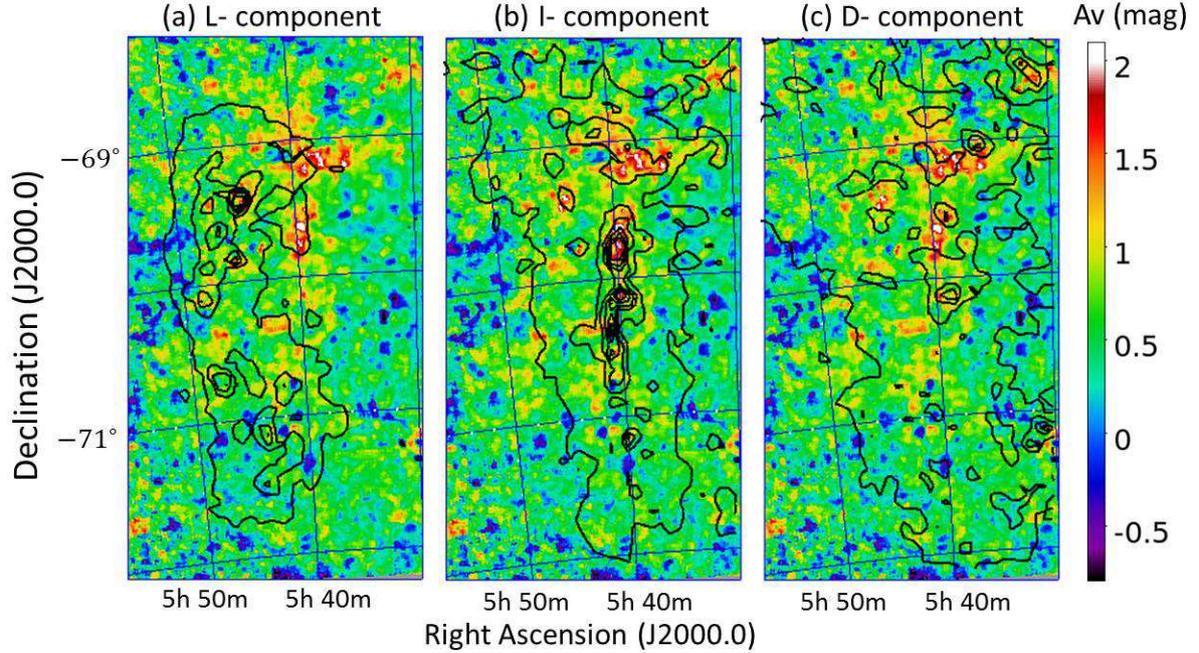} 
 \end{center}
\caption{Dust extinction map superposed on the $N(\rm H)$ maps of the (a) L-, (b) I-, and (c) D-component's contours in the H\,\emissiontype{I} ridge region.
The contour levels are (0.6, 2.1, 3.6, 5.0, 6.5, and 8.0)$\times10^{21}\ \rm cm^{-2}$.}\label{fig:compare_av_hi}
\end{figure}

\begin{figure}
 \begin{center}
  \includegraphics[width=5cm]{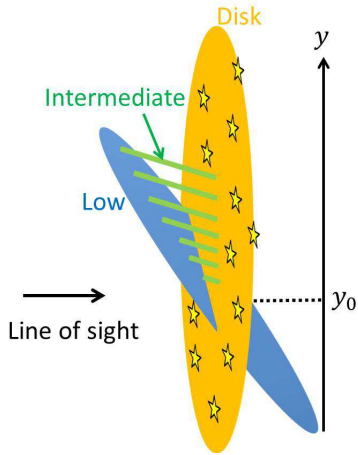} 
 \end{center}
\caption{Illustration of the expected distributions of the L-, I-, and D-components. Orange and blue disks are the gases of the LMC (D-component) and those of the origin of an inflow gas from outside the LMC (L-component), respectively. The region hatched by green lines corresponds to parts of the I-component affected by the interaction between the D- and L-components.}\label{fig:geometry}
\end{figure}

Figure \ref{fig:compare_av_hi} shows comparison of the dust extinction map whose angular resolution is reduced to be \timeform{2'.6} with the $N(\rm H)$ maps of the L-, I-, and D-component in the H\,\emissiontype{I} ridge region.
Here the mean $\delta A_V$ is improved from 0.51 mag (figure \ref{fig:extinction}b) to 0.31 mag at the sacrifice of the angular resolution, calculated by the method described in section 2.2.
\citet{fukui_2017} suggested that the L-component may have been behind the LMC disk and currently located nearly at the same position as the LMC disk, indicating on-going collisional interaction between the D- and L-components.
By comparing the dust extinction with the L-component as shown in figure \ref{fig:compare_av_hi}a, we find that the L-component spatially correlates well with the dust extinction in the northern part of the H\,\emissiontype{I} ridge region ($\delta_{\rm J2000.0} >$ \timeform{-70D}) but not in the southern part.
This trend is consistent with the suggestions by \citet{fukui_2017}.
More specifically, we assume the gas distribution illustrated in figure \ref{fig:geometry} where the L-, I- and D-components are present in the order of the near to far side from observers along the line of sight at $y > y_0$.
Here, $y_0$ indicates that the L-component at $y >y_0$ is located in front of the LMC disk, contributing to the dust extinction, whereas the L-component at $y < y_0$ is located behind the LMC disk, not contributing to the dust extinction.
To compare $A_V$ with  $N(\rm H)$ of each velocity component on a pixel-by-pixel basis, we perform linear regression with the following equation:
\begin{eqnarray}\label{eq:regress1}
A_V = \theta(y- y_0)\ a N({\rm H})_{\rm L}\ + \ b N({\rm H})_{\rm I}\ + \ c N({\rm H})_{\rm D}\ +\ C, \\ 
\theta(y- y_0) = \left\{ \begin{array}{ll}
    1 \ & (y\ge y_0), \\
    0 \ & (y < y_0),
  \end{array} \right. \nonumber
\end{eqnarray}
%
where $a,\ b,$ and$\ c$ are free parameters, coefficients proportional to the dust/gas ratios of the L-, I-, and D-components, respectively, while $N({\rm H})_{\rm L}$, $N({\rm H})_{\rm I}$, and $N({\rm H})_{\rm D}$ are the gas column densities of the L-, I-, and D-components, respectively.
The constant component $C$ is expected to account for the Galactic foreground extinction.
$\theta (y-  y_0)$ is a step function, where $y_0$ is a free parameter.
Here, as a first step, we assume that the dust/gas ratio of each component is constant over the H\,\emissiontype{I} ridge region.
Only the region where $N({\rm H})_{\rm D}$ is higher than $1.0 \times 10^{20} \ \rm cm^{-2}$ is used to perform the linear regression.
In addition, we mask the 30 Dor region where the surface brightness is higher than $20\ \rm MJy/sr$ in the Spitzer/MIPS $24\ \rm \mu m$ map from the SAGE program (\cite{mips}), because \citet{30dor} suggested that the extinction curve around the 30 Dor region does not follow the reddening law given by \citet{red_law2}.

The resultant $A_V$/$N(\rm H)$ of each velocity component is shown as model A in table \ref{tab:result}.
Here and hereafter, the uncertainties associated with the free parameters are estimated by the formal regression errors using the errors on $A_V$.
The constant component $C$ is estimated to be $0.31 \pm 0.01$ mag, while \citet{dobashi} showed that the Galactic extinction towards the LMC is $A_V \simeq 0.2$ mag.
Thus, $C$ is likely attributed to the Galactic foreground extinction.
The $\chi^2$ values divided by the degrees of freedom ($\chi^2$/dof) is 4666.03/3943, which indicates that the fit is not acceptable for a 90\% confidence level (i.e., $\chi^2$/dof $\lesssim 1.13$ is required for dof $\gtrsim 200$).
Comparing the residual extinction map after subtracting the model-predicted $A_V$ with $N$($\rm H_2$) of the I-component, we find that the region showing the negative residual coincides with the $N$($\rm H_2$) distribution of the I-component and therefore the $X_{\rm CO}$ factor of the I-component is likely to be over-estimated.
Thus, to improve the reduced $\chi^2$, we perform the linear regression with the following  equation, allowing the $X_{\rm CO}$ factor of the I-component to vary,

\begin{eqnarray}\label{eq:regress2}
A_V\ =\ \theta(y- y_0)\ a N({\rm H})_{\rm L}\ + \ b \bigl[ N({\rm H}\,\emissiontype{I})_{\rm I} + 2xW_{\rm CO} \bigr] \ + \ c N({\rm H})_{\rm D}\ +\ C, \\
\theta(y- y_0) = \left\{ \begin{array}{ll}
    1 \ & (y\ge y_0), \\
    0 \ & (y < y_0),
  \end{array} \right. \nonumber
\end{eqnarray}
where $W_{\rm CO}$ is the integrated CO intensity of the I-component, and $x$ is a free parameter corresponding to $X_{\rm CO}$ of the I-component.
In order to verify whether or not the new free parameter $X_{\rm CO}$ improves the fit significantly, we perform an F-test and calculate an F-test probability.
We adopt the threshold of the F-test probability smaller than 0.10, which indicates that the newly-introduced free parameter improves the fit for a 90\% confidence level.

\begin{figure}
 \begin{center}
  \includegraphics[width=17cm]{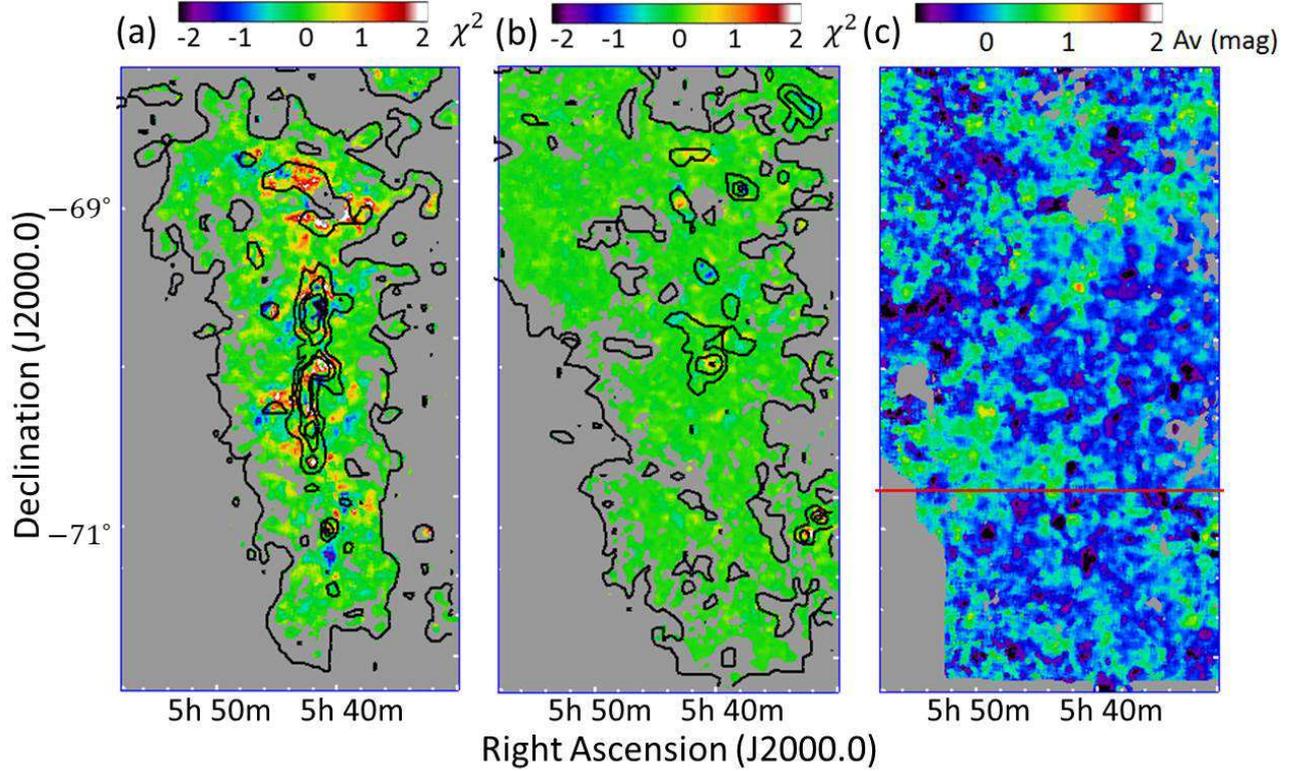} 
 \end{center}
\caption{(a) Difference in $\chi^2$ between the residual maps obtained from equations (\ref{eq:regress1}) and (\ref{eq:regress2}).
We mask the region where $N({\rm H})$ of the I-component is lower than $8.0 \times 10^{20} \ \rm cm^{-2}$.
The contours show the $N(\rm H)$ distribution of the I-component, whose levels are (0.6, 2.4, 4.2, and 6.0)$\times 10^{21}\ \rm cm^{-2}$.
(b) Difference in $\chi^2$ between the residual map obtained from equation (\ref{eq:regress2}) and that obtained from the linear regression allowing the $X_{\rm CO}$ values of the I- and D-components to vary.
The region where $N({\rm H})$ of the D-component is lower than $8.0 \times 10^{20} \ \rm cm^{-2}$ is masked.
The contours show the $N(\rm H)$ distribution of the D-component, whose levels are the same as those in panel (a).
 (c) Residual extinction map after subtracting the best-fit $A_V$ map with the linear regression allowing the $X_{\rm CO}$ values of the I- and D-components to vary. The color levels are the same as those in figure \ref{fig:extinction}. A red horizontal line is the position of the free parameter $y_0$ derived from the linear regression.
We mask the region where $N({\rm H})$ of the D-component is lower than $1.0 \times 10^{20} \ \rm cm^{-2}$ as well as the 30 Dor region.}
\label{fig:residual}
\end{figure}

As a result of the linear regression of equation (\ref{eq:regress2}), the $\chi^2$/dof value is reduced from 4666.03/3943 to 4577.53/3942; an F-test probability is $\sim 4 \times 10^{-18}$, which verifies the validity of introducing the new free parameter $X_{\rm CO}$.
The difference in $\chi^2$ between the residual maps obtained from equations (\ref{eq:regress1}) and (\ref{eq:regress2}) is shown together with the contours of $N$(H)$_{\rm I}$ in figure \ref{fig:residual}a, where we can recognize that changing $X_{\rm CO}$ of the I-component does work to improve the fitting.
The resultant $A_V$/$N(\rm H)$ and $X_{\rm CO}$ are shown as model B in table \ref{tab:result}.
Yet, the fit is still marginally acceptable for a 90\% confidence level.
Thus, we also allow the $X_{\rm CO}$ factor of the D-component to vary as well.
As a result, $\chi^2$/dof is reduced to 4566.18/3941, and F-test probability is $\sim 2 \times 10^{-3}$, again indicating the validity of introducing the new free parameter $X_{\rm CO}$ of the D-component.
Figure \ref{fig:residual}b shows the difference in $\chi^2$ between the residual maps before and after allowing the $X_{\rm CO}$ factor of the D-component to vary, together with the contours of $N$(H)$_{\rm D}$.
Here, again, we can recognize that changing $X_{\rm CO}$ of the D-component does work to improve the fitting.
The fitting results are summarized as model C in table \ref{tab:result}.
Furthermore we try the linear regression setting $X_{\rm CO}$ of the L-component to be a free parameter, however, $\chi^2$/dof and F-test probability are 4566.16/3940 and 0.89, respectively, indicating that introducing the new free parameter $X_{\rm CO}$ of the L-component is not statistically required.
The results of the linear regression are shown as model D in table \ref{tab:result}, where we can recognize that the best-fit $X_{\rm CO}$ factor of the L-component is consistent with the fixed $X_{\rm CO}$ factor.
In the following discussion, we use the results of model C in table \ref{tab:result}.
The residual map after subtracting the best-fit $A_V$ map from the observed $A_V$ map is shown in figure \ref{fig:residual}c, and the reduced $\chi^2$ is 1.16.
Hence we successfully decompose the $A_V$ distributions to those associated with the L-, I- and D-components.
The local residuals seen in figure \ref{fig:residual}c are probably caused by the foreground contamination of the D-component, which reflects that all the stars selected in our sample are not necessarily located behind the gas of the LMC disk.
\begin{table}
\tbl{Parameters derived from the comparison of the dust extinction and $N(\rm H)$ of each component.}{%
 \begin{tabular}{lccccc}
\hline
 \multicolumn{1}{c}{Model} &$\chi^2$/dof & Parameter & L-component & I-component & D-component \\
 \hline
A&4666.03/3943& $\frac{A_V}{N(\rm H)} /10^{-22}$ & $0.78\pm 0.06$ &$0.72\pm 0.05$ & $0.99\pm 0.07$ \\
 &  &$X_{\rm CO} /10^{20}$ &$7 \pm 2$\footnotemark[*] &$7 \pm 2$\footnotemark[*] &$7 \pm 2$\footnotemark[*] \\
B\footnotemark[$\dagger$]&4577.53/3942& $\frac{A_V}{N(\rm H)} /10^{-22}$ & $0.70\pm 0.06$ &$1.50\pm 0.11$ & $0.82\pm 0.08$ \\
 &  &$X_{\rm CO} /10^{20}$ &$7 \pm 2$\footnotemark[*] &$1.3\pm 0.4$&$7 \pm 2$\footnotemark[*] \\
C\footnotemark[$\ddagger$] &4566.18/3941& $\frac{A_V}{N(\rm H)} /10^{-22}$ & $0.70\pm 0.06$ &$1.35\pm 0.11$ & $1.10\pm 0.12$ \\
 & & $X_{\rm CO} /10^{20}$ &$7 \pm 2$\footnotemark[*] &$1.5\pm 0.4$&$2.9\pm 1.0$\\
D\footnotemark[$\S$] & 4566.16/3940&$\frac{A_V}{N(\rm H)} /10^{-22}$ & $0.71\pm 0.08$ &$1.35\pm 0.11$ & $1.10\pm 0.12$ \\
 & &$X_{\rm CO} /10^{20}$ &$6.8\pm 1.6$ &$1.5\pm 0.4$&$2.8\pm 1.0$\\
 \hline
 \end{tabular}}\label{tab:result}
 \begin{tabnote}
 \footnotemark[*] Fixed value.\\
\footnotemark[$\dagger$] Allowing the $X_{\rm CO}$ factor of the I-component to vary.\\
\footnotemark[$\ddagger$] Allowing the $X_{\rm CO}$ factors of the I- and D-components to vary.\\
\footnotemark[$\S$] Allowing the $X_{\rm CO}$ factors of the I-, D- and L-components to vary.
\end{tabnote}
 \end{table}

\section{Discussion}
\subsection{Dust/gas ratio}
Comparing the dust extinction map with the $N(\rm H)$ map of each velocity component, we find difference by a factor of about 2 in $A_V / N(\rm H)$ between the L- and the other components (see table \ref{tab:result}).
This result is likely to be caused by the difference in the metallicity between the L- and the other components.
The metallicity is known to be 0.3--0.5 $\rm Z_{\odot}$ in the LMC while it is 0.2 $\rm Z_{\odot}$ in the SMC (\cite{metal}).
Assuming the linear relationship between the dust abundance and the metallicity,  $A_V / N(\rm H)$ is expected to be different by a factor of about 2 between the LMC and the SMC.
\citet{smc_lmc} find $A_V / N(\rm H)= 1.39 \times 10^{22} \ \rm mag \ cm^2$ for the LMC and $0.60 \times 10^{22} \ \rm mag \ cm^2$ for the SMC from comparison between the total hydrogen column density and the visual extinction derived from Herschel dust continuum maps at 160 $\rm \mu m$.
In our result, the $A_V / N(\rm H)$ values of the I- and L-components are consistent with those of the LMC and the SMC, respectively.
$A_V / N(\rm H)$ of the D-component agrees with that of the I-component within the uncertainties.
Therefore, our result supports the scenario that the L-component is of the origin of an inflow gas from the SMC and is possibly mixed with the gas in the LMC as suggested by \citet{fukui_2017}.

\citet{fukui_2017} compared the H\,\emissiontype{I} intensity, $W$(H\,\emissiontype{I}), with the dust optical depth at 353 GHz, $\tau_{353}$. 
The integrated velocity range of $W$(H\,\emissiontype{I}) is $V_{\rm offset}\ =$ $-$100 to 90 $\rm km\ s^{-1}$ covering all the velocity ranges of the L-, I-, and D-components.
They found that $W(\rm H\,\emissiontype{I}) / \tau_{353}$ in the H\,\emissiontype{I} ridge region is about 2 times higher than that in the stellar bar region, and concluded that the gases in the LMC H\,\emissiontype{I} ridge region are contaminated with an inflow gas from the SMC. 
It should be noted that we for the first time evaluate the dust/gas ratios of the L-, I-, and D-components separately in the H\,\emissiontype{I} ridge region.
In our result, $A_V / N(\rm H)$ of the L-component is 2 times lower than that of the other components.
Thus, we clearly demonstrate that the low metallicity gas is actually present in the H\,\emissiontype{I} ridge region.

In order to estimate the effect of the uncertainties of $X_{\rm CO}$, we mask the region in our dust extinction map where $W_{\rm CO} > 1.2$ $\rm{K\ km\ s^{-1}}$ corresponding to the $3\sigma$ noise level of the CO detection (\cite{h2_conv}), and perform the linear regression using equation (\ref{eq:regress1}) without introducing $X_{\rm CO}$ factors.
As a result, the $A_V / N(\rm H)$ values of the L-, I-, and D-components are estimated to be ($0.76\pm 0.07$, $1.16\pm 0.08$, and $0.94\pm 0.09$)$\times 10^{-22}$, respectively, 
which are consistent with those of model C in table \ref{tab:result} within the errors.
Thus, introducing the $X_{\rm CO}$ factors does not affect the above conclusion.

\subsection{CO-to-H$_2$ conversion factor}
We also find difference in the $X_{\rm CO}$ factors between the L- and the other components (see table \ref{tab:result}).
$X_{\rm CO}$ is known to be dependent on the metallicity (\cite{xco_metal}).
In the low-metallicity environments, CO is photodissociated by ultraviolet photons due to the lack of dust shielding.
Thus, less CO emission traces the $\rm H_2$ column density, leading to a higher $X_{\rm CO}$ factor.
Our result suggests that the $X_{\rm CO}$ factor of the L-component is higher than those of the other components, which implies that the gas of the L-component has lower metallicity.
However, the $X_{\rm CO}$ factor of the D-component (i.e, the LMC disk) derived from our analysis is smaller than those derived from a virial analysis (e.g., $X_{\rm CO} \sim 7 \times 10^{20}$, \cite{h2_conv}; $X_{\rm CO} \sim 4 \times 10^{20}$, \cite{co_1}; \cite{co_2}).
As mentioned above, CO is photodissociated in low-metallicity environments, which results in a higher $X_{\rm CO}$ factor.
\citet{smc_lmc} correct this effect, and show that the $X_{\rm CO}$ factors in the LMC and the SMC are $2.9\times10^{20}$ and $7.6\times10^{20}\ {\rm cm^{-2}}/(\rm{K\ km\ s^{-1}})$, respectively.
The $X_{\rm CO}$ factor of the D-component is consistent with that of the LMC.
Thus, the discrepancy between the extinction-based and viral mass-based estimates of $X_{\rm CO}$ is possibly caused by the photodissociation of CO.
The $X_{\rm CO}$ factor of the L-component is similar to that of the SMC, which is also consistent with the scenario that the L-component is of the origin of an inflow gas from the SMC. 

\subsection{Geometry of the gas}
We obtain the best-fit free parameter $y_0$ of $147 \pm 25$ pix corresponding to $\delta_{\rm J2000.0} =$ \timeform{-70D.8}$\pm$\timeform{0D.2} at $\alpha_{\rm J2000.0} = $\timeform{87D.4}.
This implies that the L-component is located in front of the LMC disk above the boundary denoted by the red horizontal line in figure \ref{fig:residual}c.
We evaluate the appropriateness of this boundary position, referring to the position-velocity diagram in figure 2 of \citet{fukui_2017}, where the integrated range in R.A. is from \timeform{86D.69} to \timeform{87D.41}.  
In their position-velocity diagram, the several bridge features can be seen at  $\delta_{\rm J2000.0} =$ \timeform{-68D.8} to \timeform{-70D.5}, which are the evidence for interactions between the L- and D-components.
Our boundary position of $\delta_{\rm J2000.0} =$ \timeform{-70D.8}$\pm$\timeform{0D.2} is consistent with the position where the bridge features appear, which supports the collision between the L- and D-components.
Considering the illustrative geometry of the gas shown in figure \ref{fig:geometry}, the gases of the L- and I-components are expected to be located in front of the stars in the LMC disk.
Therefore, we can evaluate $A_V$/$N$(H) of both components reliably, whereas $A_V$/$N$(H) of the D-component has the uncertainties caused by the positional relationship between the stars in the LMC disk and the gas of the D-component.

As a whole, we for the first time evaluate $A_V$/$N$(H) of the L-, I- and D-components separately in the H\,\emissiontype{I} ridge region, and in particular estimate $A_V$/$N$(H) of the L- and I-components reliably based on the 3-D geometry of the gas expected from the velocity-resolved observations of \citet{fukui_2017} and the comparison of $A_V$ and $N$(H) in this paper.
Both $A_V$/$N$(H) and $X_{\rm CO}$ of the L-component consistently suggest that the gas of the L-component is of the origin of an inflow gas from the SMC.

\section{Conclusion}
We create a new dust extinction map of the LMC in the H\,\emissiontype{I} ridge region,  using the IRSF data with the updated calibration.
We compare the dust extinction with the multiple cloud components of different velocities recently revealed by \citet{fukui_2017}, and evaluate the dust/gas ratio, $A_{V}$/$N$(H), of the different velocity components.
Our main results are as follows:
\begin{enumerate}
\item the dust extinction map derived from the near-IR color excess correlates well with the $N$(H) map obtained from the CO and H\,\emissiontype{I} observations. 
The spatial resolution of our IRSF extinction map is improved by a factor of 2, as compared to the previous extinction map from the 2MASS near-IR color excess presented by \citet{dobashi}.
\item $A_{V}$/$N$(H) of the L-component is significantly lower than those of the other velocity components and is consistent with that of the SMC, while $A_{V}$/$N$(H) of the I- and D-components are consistent with that of the LMC (\cite{smc_lmc}).
This result suggests that the low-metallicity gas from the SMC may be contaminated in the LMC H\,\emissiontype{I} ridge region, which supports the numerical simulation by \citet{shock_sim} and is consistent with the observational fact that $W$(H\,\emissiontype{I})/$\tau_{353}$ in the H\,\emissiontype{I} ridge region is higher than that outside the H\,\emissiontype{I} ridge region (\cite{fukui_2017}). 
\item The $X_{\rm CO}$ factor of the L-component is higher than those of the other components and is similar to that of the SMC, while the $X_{\rm CO}$ factors of the I- and D-components are consistent with that of the LMC (\cite{smc_lmc}).
This result also favors the scenario that the L-component originates from the low-metallicity gas from the SMC.

\end{enumerate}
As a whole, our results are likely to support the scenario that the gas in the H\,\emissiontype{I} ridge region is contaminated with an inflow gas from the SMC with a geometry consistent with the on-going collision between the two velocity cloud components.
\begin{ack}
We thank the referee for giving us many useful comments.
We also thank Prof. Kazuhito Dobashi for kindly providing us with the electronic data of their $A_V$ map.
The IRSF project is a collaboration between Nagoya University
and the SAAO supported by the Grants-in-Aid for Scientific Research on Priority Areas (A) (Nos. 10147207 and 10147214) and Optical \& Near-Infrared Astronomy Inter-
University Cooperation Program, from the Ministry of Education,
Culture, Sports, Science and Technology (MEXT) of Japan and the National Research Foundation (NRF) of South Africa.
\end{ack}





\end{document}